\documentclass{amsart}

\usepackage{amssymb}
\usepackage{amsmath}
\usepackage{graphicx}

\begin{document}

\title{The construction of variance estimators for particulate material sampling}

\author{Bastiaan Geelhoed}
\address{Delft University of Technology, Mekelweg 15, 2629 JB, Delft, The Netherlands}
\email{b.geelhoed@tudelft.nl}
\urladdr{http://homepage.tudelft.nl/4h8k2}

\begin{abstract}
The variance of the concentration in a sample can be estimated using knowledge of the particle masses, concentrations and the parameter for the dependent selection of particles. A number of variance estimators are constructed including a class of hybrid estimators.
\end{abstract}

\maketitle

\section{Introduction}
\label{intro}
Particulate materials are routinely sampled in industries that deal with solid materials and generally the
concentration of a certain substance of interest in a sample is used as an estimate for the corresponding
concentration in the population (or batch) from which the sample was taken.
Knowledge of the occurrence of sampling errors, made during this estimation, and their
potential magnitude is a prerequisite to being able to make reliable decisions based on samples.
The terms "population" and "batch" are used as synonyms as the term "batch"
is commonly used in the application area of particulate material sampling. Particulate materials
are materials that
consist of solid objects, called "particles".
The size of these particles can range from the scale of a micrometer or smaller for powders to the scale of
a centimeter or larger for coarse granular materials. 

	Our framework can mathematically be formulated as follows: there is a large but finite population
made of $T$ kinds of particles. We will let $m_i$ denote the mass of one particle of the
$i$th kind and we will
let $c_i$ denote the (mass) concentration of the substance of interest in a particle of kind $i$. Both $m_i$ and $c_i$ are assumed known, or can be accurately
measured. The population is sampled (according to a sampling design), which implies here that a single
sample is obtained which contains a number of particles. The number of particles of the $i$th kind in the sample
are counted and recorded as $N_i$. Note
that the sample size $\sum _i N_i$ is a random variable and not assumed to be fixed here. The total mass of the
sample is $M_{sample}$ defined by:
\begin{eqnarray}
   M_{sample}  =  \sum_{i=1}^{T} N_i m_i  \nonumber
\end{eqnarray}
and the total mass of the substance of interest in this sample is $A_{sample}$ defined by:
\begin{eqnarray}
   A_{sample}  =  \sum_{i=1}^{T} N_i m_i c_i  \nonumber
\end{eqnarray}
The sample concentration (denoted by $\hat{\theta}$) is defined by as $\hat{\theta} = A_{sample}/M_{sample}$ and is
an estimator for the corresponding population quantity
denoted as $\theta$, i.e. the ratio of the amount
of property of interest in the batch $A_{batch}$ and the mass of the batch $M_{batch}$.
This article discusses how to estimate the variance of $\hat{\theta}$. A general approach to the problem of finding variance estimators for $\hat{\theta}$ is followed
which attempts to construct
variance estimators that are applicable for a wide range of possible sampling designs.
	
	In order to specify estimators for the variance, without specifying an explicit sampling design,
we need some information concerning the first and second-order inclusion probabilities, which are assumed
to be well-defined even though the sampling design remains unspecified here. For this purpose, it will be seen
that a new parameter, "the parameter for the dependent selection of particles" (denoted by $C_{ij}$ and
discussed in section \ref{dep_sel_of_part} below), suffices. Like $m_i$, $c_i$ and $N_i$, it will be
assumed $C_{ij}$ is known or can be accurately determined. Practical determination of the required
parameters ($N_i$, $m_i$, $c_i$ and $C_{ij}$) is discussed in section \ref{prac_det}.

	The variance of $\hat{\theta}$ is both influenced by variations in $M_{sample}$ and in $A_{sample}$. 
Because $M_{sample}$ can potentially vary between the mass of the lightest particle in the population and the mass
of the population and $A_{sample}$ can potentially vary between zero and the total amount of property of interest
in the population, the variance of $\hat{\theta}$  remains finite. A variance estimate can therefore serve as a
practically useful quantity to gain insight into the potential magnitude of the sampling error.
In this article, a number of estimators  for the variance of
$\hat{\theta}$ will be constructed. For minimization of the bias of $\hat{\theta}$ the reader is referred to other literature
(see e.g. \cite{r12}).

\section{Dependent selection of particles}
\label{dep_sel_of_part}
It will be seen that the variance of $\hat{\theta}$ depends on the particle masses (the values of $m_i$),
the particle concentrations (the values of $c_i$), and also on the covariance matrix of the variables
$N_1, ..., N_T$ via the "parameter for the dependent selection of particles" ($C_{ij}$), defined by
Eq. (\ref{Cij}) below. In that expression, the effect of dependent selection of particles on the covariance
matrix of the variables $N_1, ..., N_T$ is parameterized as:
\begin{equation}
\label{Cij}
E\left(N_i N_j \right)- E\left(N_i \right) E\left(N_j \right)  = \Delta _{ij} E\left(N_i \right) - C_{ij} E\left(N_i \right) E\left(N_j \right)
\end{equation}
Where $E\left( . \right)$ denotes an expected value and $\Delta_{ij}$ is the Kronecker delta which is one
when $i=j$ and zero otherwise. Note that Eq. (\ref{Cij}) implies that each $N_i$ does not have to have a
marginal Poisson distribution, because the matrix with elements $C_{ij}$ may have non-zero diagonal elements.
In addition, $N_i$ and $N_j$ may be correlated, because the off-diagonal elements of the matrix with elements
$C_{ij}$ may also be non-zero. In other words, non-zero $C_{ii}$ are used to parameterize deviations from the
marginal Poisson distribution, while non-zero values of $C_{ij}$ for $i \neq j$ are used to parameterize deviations
from zero correlation between $N_i$ and $N_j$.

It can be proven -under certain conditions (see \ref{app})- that $C_{ij}$ can be interpreted as a correction
for the dependent selection of particles in the limit of a sufficiently large population. Denoting the inclusion
probability of a particle belonging to the $i$th and $j$th particle class respectively as $\kappa _i$ and
$\kappa _j$, and denoting the second-order inclusion probability of the pair consisting of a particle of type
$i$ and $j$ as $\kappa _{ij}$, this can be expressed as:
\begin{equation}
\label{Cij_inc}
C_{ij}  \approx 1 - \frac{\kappa _{ij}}{\kappa _i \kappa _j}
\end{equation}
In \ref{app}, a derivation is presented which shows that the '$\approx$' in the above equation can be
replaced by '$=$' when the population (or batch) contains an infinite number of particles of each type or kind.
Apart from a sign, the above equation corresponds to the result given by \cite{r3}. From Eq. (\ref{Cij_inc})
follows that when $C_{ij}=0$ the particle selections can be considered approximately independent
($\kappa _{ij} \approx \kappa _i \kappa _j$). It also follows from Eq. (\ref{Cij_inc}) that if $C_{ij}$ is
positive, $\kappa _{ij} < \kappa _i \kappa _j$ , which can be caused by segregation of particles of type
$i$ and $j$. If $C_{ij}$ is negative, $\kappa _{ij} > \kappa _i \kappa _j$ , which can be caused by grouping
of particles of type $i$ and $j$. Thus, the parameter $C_{ij}$ has a physical meaning, which makes the parameterization
expressed in  Eq. (\ref{Cij}) practically significant.

\section{Construction of variance estimators}
\label{construction}

\subsection{Estimators based on a Taylor expansion}

\subsubsection{Linear terms (i)}
A Taylor-linearization (see e.g. chapter 5 of \cite{r5}) of $\hat{\theta} = A_{sample}/M_{sample}$ with respect
to deviations of $A_{sample}$ and $M_{sample}$ from their expected values is applied:
\begin{eqnarray}
\hat{\theta} = \frac{A_{sample}}{M_{sample}} \doteq 
\nonumber \\
\frac{E\left( A_{sample} \right)}{E\left( M_{sample} \right)} + \frac{1}{E\left( M_{sample} \right)} \left( A_{sample} - E\left( A_{sample} \right) \right)
+ \nonumber \\ 
 \frac{E\left( A_{sample} \right)}{E\left( M_{sample} \right)} \left[ -\frac{1}{E\left( M_{sample} \right)} \left( M_{sample} - E\left( M_{sample} \right) \right) \right]
 \nonumber
\end{eqnarray}
where $\doteq$ means "is equal to in first-order Taylor series expansion".  From the above linearization it follows that the variance of $\hat{\theta}$  is approximated by:
\begin{eqnarray}
V(\hat{\theta})\doteq
\nonumber \\
\frac{
 V(A_{sample}) + \frac{E^2(A_{sample})}{E^2(M_{sample})}V(M_{sample}) -2
 \frac{E(A_{sample})}{E(M_{sample})}Cov(A_{sample};M_{sample}) 
}{E^2(M_{sample})}
\nonumber
\end{eqnarray} 
Where $V(.)$ and $Cov(.;.)$ represent the variance and covariance operator respectively. Expressing the expected
values, variances and covariance in terms of the particle numbers $N_i$ and $N_j$ yields:

\begin{eqnarray}
\label{T1}
 V(\hat{\theta})\doteq - \frac{1}{E^2(M_{sample})} \times
\nonumber \\
 \sum _{i=1}^T \sum _{j=1}^T m_i m_j \left( c_i - \frac{E(A_{sample})}{E(M_{sample})}\right) \left( c_j - \frac{E(A_{sample})}{E(M_{sample})} \right)
 \nonumber \\
 \times \left[ E(N_i N_j) - E(N_i)E(N_j) \right] 
\end{eqnarray}

Substituting Eq. (\ref{Cij}) into Eq. (\ref{T1}) results in:

\begin{eqnarray}
V(\hat{\theta})\doteq \frac{1}{E^2(M_{sample})} \sum _{i=1}^T E(N_i) m_i^2\left( c_i - \frac{E(A_{sample})}{E(M_{sample})}\right)^2
- \frac{1}{E^2(M_{sample})} \times
\nonumber \\
 \sum _{i=1}^T \sum _{j=1}^T m_i m_j \left( c_i - \frac{E(A_{sample})}{E(M_{sample})}\right) \left( c_j - \frac{E(A_{sample})}{E(M_{sample})} \right) C_{ij} E(N_i)E(N_j) 
\nonumber
\end{eqnarray}
The above expression is used to derive an expression for a variance estimator by replacing
$E(N_i)$ by $N_i$ and $E(N_j)$ by $N_j$. Regrouping terms slightly yields:
\begin{eqnarray}
\label{VT1}
 \widehat{V}_{T1}(\hat{\theta})=  \frac{1}{M^2_{sample}}  \sum _{i=1}^T \sum _{j=1}^T m_i m_j \left( c_i - \hat{\theta} \right) \left(  c_j - \hat{\theta}\right)\left[ N_i \Delta _{ij} - C_{ij} N_i N_j \right]
\end{eqnarray}
The index $T1$ is used to denote the above estimator ("the variance estimator based on a first-order Taylor expansion"),
because of the first-order Taylor expansion. The estimator was also given by \cite{r3}.

\subsubsection{Second-order terms (ii)}
Although it is expected that the first-order Taylor series expansion will usually work well, it is interesting to
construct a novel estimator based on the second-order Taylor expansion. The second-order Taylor expansion of
$\hat{\theta}$ is:
\begin{eqnarray}
\label{T2}
\hat{\theta} = \frac{A_{sample}}{M_{sample}} \doteq cst + 
\frac{E\left( A_{sample} \right)}{E\left( M_{sample} \right)} 
\times \nonumber \\
\left[  \frac{2A_{sample}}{E(A_{sample})} - \frac{2M_{sample}}{E(M_{sample})} 
- \frac{A_{sample}M_{sample}}{E(A_{sample})E(M_{sample})}
+ \frac{M^2_{sample}}{E^2(M_{sample})}
 \right ]
\end{eqnarray}
where $\doteq$ means "is equal to in second-order Taylor series expansion" and
$cst$ represents some terms that are constant and which therefore do not influence the variance of
$A_{sample}/M_{sample}$. Calculating the variance using the above equation directly would imply evaluation
of terms like $E(A^x_{sample} M^y_{sample})$ with $x+y=3$ or $x+y=4$, for which Eq. (\ref{Cij}) is no help.
To be able to use Eq. (\ref{T2}), further assumptions are therefore required. A statistical model is proposed
and used here, which states that a new quantity, $B_{sample}$, defined as:
\begin{equation}
\label{B_sample}
B_{sample} = A_{sample} - M_{sample} Cov(A_{sample} ; M_{sample})/V(M_{sample})
\end{equation}
is independent on $M_{sample}$. This model leads to a zero covariance between $B_{sample}$ and $M_{sample}$, while
preserving the covariance between $A_{sample}$ and $M_{sample}$.  It is also assumed that $M_{sample}$ has the
same skewness and kurtosis as a normal distribution (this assumption is discussed below).
Using the here assumed statistical model, assumption and
Eq. (\ref{T2}) results after a lengthy computation in:

\begin{eqnarray}
\label{VB}
V(\hat{\theta}) \doteq \frac{V(B_{sample})}{E^2(M_{sample})} +
\frac{  V(M_{sample})   }
{E^4\left( M_{sample} \right)}
\times \nonumber \\ 
 \left[  E^2(B_{sample}) +  V(B_{sample}) +2 \beta ^2 \frac{E^2(A_{sample})}{E^2(M_{sample} )} V(M_{sample}) \right]
\end{eqnarray}

where $\beta$ is defined by:
\begin{eqnarray}
\label{beta}
\beta = 1 - \frac{ Cov(A_{sample}; M_{sample}) E(M_{sample}) } { V(M_{sample}) E(A_{sample}) }
\end{eqnarray}
The variable $B_{sample}$ can be eliminated in the Eq. (\ref{VB}) using Eq. (\ref{B_sample}). The resulting expression
depends on the expected values, variances and covariances of $A_{sample}$ and $M_{sample}$. In this expression,
$A_{sample}$ and $M_{sample}$ can be written in terms of the variable $N_i$. Eq. (\ref{Cij}) can subsequently be
used to eliminate the covariances between $N_j$ and $N_j$ on the right-hand side of the expression. The result is an
expression that depends on $m_i$, $c_i$, $E(N_i)$ and $C_{ij}$. The expected values $E(N_i)$ can then be replaced by
their sample values $N_i$ to obtain a variance estimator. This estimator (not fully written out here because it is a
long equation) is named "the variance estimator based on a second-order Taylor expansion" and is denoted by the
symbol $\widehat{V}_{T2}\left( \hat{\theta} \right)$ , where the index $T2$ refers to the second-order Taylor expansion.

When the assumptions with respect to skewness and kurtosis are not met, the estimator
$\widehat{V}_{T2}\left( \hat{\theta} \right)$ can have an increased bias. It is expected, however,
that in many practical scenarios the sample mass is normally distributed - an assumption that can be tested.
If the distribution of sample masses is non-normal, the estimator $\widehat{V}_{T2}\left( \hat{\theta} \right)$ can
possibly be adapted to take into account the skewness and excess kurtosis of the distribution of $M_{sample}$.

\subsection{A variance estimator based on the Horvitz-Thompson estimator}
The Horvitz-Thompson estimator for the variance of a $\pi$-expanded estimator can used to find an estimator for
the variance of $\hat{\theta}$. A $\pi$-expanded estimator (see e.g. \cite{r5}) for the concentration in the batch
can be obtained by considering the batch concentration ($\theta = A_{batch}/M_{batch}$) to be the population total
of $y_i=m_{n(i)} c_{n(i)}/M_{batch}$, where $m_x$ and $c_x$ denote the mass and concentration of a particle of type
$x$ (for $x$ between one and the total number of classes), $n(i)$ denotes the class of the $i$th particle in the
batch and $M_{batch}$ is the total mass of the population (or batch). The expression is:
\begin{equation}
\label{pi_est}
\hat{\theta}_{\pi} = \sum _{i=1} ^T N_i m_i c_i /(M_{batch}\kappa _i)
\end{equation}
where $\hat{\theta}_{\pi}$ is the $\pi$-expanded estimator for the concentration in the batch, $N_i$ is the number
of particles in the sample belonging to the $i$th particle class, $m_i$ and $c_i$ are respectively the mass of and
the concentration in a particle belonging to the $i$th class, $M_{batch}$ is the mass of the batch and $\kappa _i$
is defined here as the first-order inclusion probability of a particle belonging to the $i$th class. (In this article two symbols ($\kappa$ and $\pi$) are used to denote inclusion probabilities, which have a subtle difference in meaning of the
index: for $\kappa$ the index refers to the kind of the particle, while for $\pi$ the index refers to the particle number.
Hence, by the definition adopted in this article: $\kappa _i = \pi _j$ when particle $j$ of the population is of kind $i$. Implicit in the use of the variable 
$\kappa _i$ is the assumption that $\pi _i$ does not have variation between particles of the same kind.)

A derivation of Eq. (\ref{pi_est})
was given by \cite{r1}. If the sample mass is constant and the first-order inclusion probability is equal
to the ratio of the sample mass ($M_{sample}$) and the batch mass ($M_{batch}$), the $\pi$-expanded estimator
becomes equal to the sample concentration, $\hat{\theta}$. Under these assumptions, the following equation for
the variance of the sample concentration, based on the general Horvitz-Thompson estimator for the variance of
the $\pi$-expanded estimator, can be derived:
\begin{equation}
\widehat{V}_{HT}(\hat{\theta})= \sum _{i=1} ^T  \sum _{j=1} ^T N_i N_j 
\left(  \frac{1}{\kappa _i \kappa_j} - \frac{1}{\kappa _{ij}} \right)
\frac{m_i m_j c_i c_j}{M^2_{batch} }
+  \sum _{i=1} ^T N_i \left( \frac{1}{\kappa _{ii}} - \frac{1}{\kappa _i} \right) \frac{m^2_i c^2_i}{M^2_{batch}}
\end{equation}
in which $\kappa _{ij}$ is the second-order inclusion probability of a particle pair in which the first-particle
belongs to the $i$th class and the second to the $j$th class. (i.e. a similar relation between 
$\kappa _{ij}$  and $\pi _{ij}$ exists as between $\kappa _{i}$  and $\pi _{i}$  ). Substitution of
Eq. (\ref{Cij_inc}) (i.e. $\kappa _{ij} = \kappa _i \kappa _j (1-C_{ij})$) for the second-order inclusion
probability and $\kappa _i = M_{sample}/M_{batch}$, results in:
\begin{eqnarray}
\widehat{V}_{HT}(\hat{\theta})= \sum _{i=1} ^T  \sum _{j=1} ^T N_i N_j 
\left(  1 - \frac{1}{1 - C_{ij} } \right)
\frac{m_i m_j c_i c_j}{M^2_{sample} }
\nonumber \\
+  
\sum _{i=1} ^T N_i  
\left( \frac{1}{1 - C_{ii} } - \frac{M_{sample}}{ M_{batch} } \right)
\frac{m^2_i c^2_i}{M^2_{sample}}
\end{eqnarray}
Assuming that the batch (or population) from which the sample was drawn is much larger than the sample,
i.e. $M_{batch} >> M_{sample}$ so that $1/(1 - C_{ii})-M_{sample}/M_{batch} \approx 1/(1 - C_{ii})$, the
above result can be rearranged to yield:
\begin{equation}
\label{VHT}
\widehat{V}_{HT}(\hat{\theta})= 
\frac{1}{M^2_{sample}}
\sum _{i=1} ^T  \sum _{j=1} ^T 
\frac{\left[ N_i \Delta _{ij} - C_{ij} N_i N_j \right] m_i m_j c_i c_j}{1-C_{ij} }
\end{equation}
The above estimator is named 'the variance estimator based on the Horvitz-Thompson estimator'  and denoted by the
symbol $\widehat{V}_{HT}\left( \hat{\theta} \right)$, where the index 'HT' refers to Horvitz-Thompson.
The estimator expressed in Eq. (\ref{VHT}) was also given by \cite{r3}.

Using Eq. (\ref{Cij}), it can be proven that, as expected, the above estimator is unbiased when the sample
mass $M_{sample}$ is a constant. In practice there will almost always be slight random variations in sample mass,
leading to a potential bias in the estimator. It is expected, though, that reducing variations in sample mass (e.g.
by using a sampling tool that results in samples of constant sample mass, will also reduce the possible bias caused
by variations in sample mass.

\subsection{Adaptations of variance estimators}
In Eq. (\ref{VHT}), the factor $[ N_i \Delta _{ij} - $ $C_{ij} N_i N_j ]$ $/(1-C_{ij})$ is noticeable. The question therefore rises whether the variance estimator expressed in (4) might be improved by replacing the factor $ [ N_i \Delta _{ij} - $ $C_{ij} N_i N_j ]$ by $[ N_i \Delta _{ij} - $ $C_{ij} N_i N_j ]$ $/(1-C_{ij})$, based on an analogy with $\widehat{V}_{HT}(\hat{\theta})$. This results in the following variance estimator:
\begin{eqnarray}
\widehat{V}_{AD1}(\hat{\theta})= 
\frac{1}{M^2_{sample}}
\sum _{i=1} ^T  \sum _{j=1} ^T 
\frac{\left[ N_i \Delta _{ij} - C_{ij} N_i N_j \right] m_i m_j
(c_i - \hat{\theta} )
(c_j - \hat{\theta})
}{1-C_{ij} }
\end{eqnarray}
Where the index $AD1$ refers to 'first adaptation'. Using Eq. (\ref{Cij}), it can be proven that the expected value of the above estimator is equal to the right-hand side of Eq. (\ref{T1}) if statistical fluctuations in $M_{sample}$ and $\hat{\theta}$ are discarded.
The above estimator can also be considered to be a derivation of the estimator based on the Horvitz-Thompson estimator, where $c_i$ and $c_j$ are replaced by $(c_i - \hat{\theta})$ and $(c_j - \hat{\theta})$  respectively.

A second adaptation is obtained by replacing the factors  $ [ N_i \Delta _{ij}$ $- C_{ij} N_i N_j ]$ by $[ N_i \Delta _{ij}$ $- C_{ij} N_i N_j ]/(1-C_{ij})$ in the equation for $\widehat{V}_{T2}\left( \hat{\theta} \right)$ . The resulting estimator is denoted by the symbol $\widehat{V}_{AD2}\left( \hat{\theta} \right)$ .

\subsection{A variance estimator based on the Sen-Yates-Grundy variance estimator}

Sen \cite{r6} and Yates and Grundy \cite{r7} derived the following estimator for the variance of the
$\pi$-expanded estimator for the population total of $y_i$:
\begin{eqnarray}
\widehat{V}_{SYG}= 
\frac{1}{2}
\sum _{k \in S}  \sum _{
l \in S, \; l \neq  k} 
\left(
\frac{y_k}{\pi _k}-\frac{y_l}{\pi _l}
 \right ) ^2
\frac{\pi _k \pi _l - \pi _{kl}}{\pi _{kl}}  
\nonumber
\end{eqnarray}
where $\pi _k$, $\pi _l$ and $\pi _{kl}$ denote respectively the inclusion probability of the $k$th and $l$th
particle of the population and the second-order inclusion probability of a pair consisting of the
$k$th and $l$th particle. For fixed size samples and positive second-order inclusion probabilities,
the above estimator is unbiased. Note that the indices $k$ and $l$ refer to a particle number in the population.

Applying the equation to the $\pi$-expanded estimator for the batch concentration (see Eq. (\ref{pi_est})) and
using $\pi _i = \kappa _{n(i)}$ (Eq. (\ref{kappa_i})),  $\pi _{ij} = \kappa _{n(i)n(j)}$ (Eq. (\ref{kappa_ij}))
and $\kappa _{ij}= \kappa _i \kappa _j (1 - C_{ij})$ (Eq. (\ref{Cij_inc})) yields:
\begin{eqnarray}
\widehat{V}_{SYG} (\hat{\theta} )= 
\frac{1}{2}
\sum _{k \in S} \sum _{
l \in S, \; l \neq  k}
\left(
\frac{m_{ n(k)} c_{n(k)} }{M_{batch} \kappa _{n(k))}} -
\frac{m_{ n(l)} c_{n(l)} }{M_{batch} \kappa _{n(l))}}
 \right ) ^2
\frac{C_{n(k)n(l)}}{1 - C_{n(k)n(l)}} 
\nonumber
\end{eqnarray}
Substituting $\kappa _{n(i)} =M_{sample}/M_{batch}$ results in:
\begin{eqnarray}
\widehat{V}_{SYG} (\hat{\theta} )= 
\frac{1}{2 M^2_{sample}}
\sum _{k \in S} \sum _{
l \in S, \; l \neq  k}
\left(
m_{ n(k)} c_{n(k)} -
m_{ n(l)} c_{n(l)}
 \right ) ^2
\frac{C_{n(k)n(l)}}{1 - C_{n(k)n(l)}}
\nonumber
\end{eqnarray}
The summations can be rewritten as summations over the particle classes:
\begin{equation}
\widehat{V}_{SYG} (\hat{\theta} )= 
\frac{1}{2 M^2_{sample}}
\sum _{i=1} ^T \sum _{j=1} ^T
N_i N_j (m_i c_i - m_j c_j)^2 
\frac{C_{ij}}{1 - C_{ij}} 
\end{equation}
The above estimator is named 'the estimator based on the Sen-Yates-Grundy estimator' and is denoted by the symbol
$ \widehat{V}_{SYG} (\hat{\theta} ) $, where the index $SYG$ refers to 'Sen-Yates-Grundy'.

\section{Hybrid variance estimators}
Hybrid forms, which are combinations of the estimators constructed in section \ref{construction}, can be constructed in order to combine the strengths of the estimators developed so far. Here a class of hybrid estimators is derived based on combining the variance estimator based on the first-order Taylor linearization, $\widehat{V}_{T1}(\hat{\theta})$, with the variance estimator based on the Horvitz-Thompson estimator, $\widehat{V}_{HT}(\hat{\theta})$. As noted above, $\widehat{V}_{HT}(\hat{\theta})$  is unbiased when the variance of $M_{sample}$ is zero. On the other hand, the variance estimator  $\widehat{V}_{T1}(\hat{\theta})$ is designed to take into account linear variations in $M_{sample}$. A possible suitable hybrid form would therefore be:
\begin{eqnarray}
\widehat{V}_{HYB} \left( \hat{\theta} \right) = 
a \widehat{V}_{T1}(\hat{\theta})
+ (1-a) \widehat{V}_{HT}(\hat{\theta})
\nonumber
\end{eqnarray}
Where $a$ would ideally be a monotonic increasing function of $V(M_{sample})$, with $a=0$ when $V(M_{sample})=0$ and $0<a<1$ when $V(M_{sample})>0$. Combining the definition of $M_{sample}$ and Eq. (\ref{Cij}) results in:
\begin{eqnarray}
V(M_{sample}) = \sum _i E(N_i) m ^2 _i - 
\sum _i \sum _j C_{ij} E(N_i N_j ) m_i m_j
\nonumber
\end{eqnarray}
An unbiased estimator for the variance of $M_{sample}$ is therefore:
\begin{eqnarray}
\widehat{V}(M_{sample}) = \sum _i N_i m ^2 _i - 
\sum _i \sum _j C_{ij} N_i N_j m_i m_j
\nonumber
\end{eqnarray}
The Relative Standard Deviation ($RSD$) of $M_{sample}$ can therefore be estimated using:
\begin{eqnarray}
\widehat{RSD}(M_{sample}) =
\begin{array}{cr}
\frac{\sqrt{\widehat{V}(M_{sample})}}{\sum _i N_i m_i} 
&
\mbox{if $\widehat{V}(M_{sample}) \geq 0$}
\\
 0
&
\mbox{if $\widehat{V}(M_{sample}) < 0$}
\end{array}
\nonumber
\end{eqnarray}
A flexible choice for a therefore would be:
\begin{eqnarray}
a = 1 - e^{-\widehat{RSD}(M_{sample})/x}
\nonumber
\end{eqnarray}
in which a numerical value can be assigned to $x$. The resulting variance estimator is:
\begin{eqnarray}
\widehat{V} _{HYBX}(\hat{\theta}) 
=
a     \widehat{V} _{T1} (\hat{\theta}) +
(1-a) \widehat{V} _{HT} (\hat{\theta})
\nonumber
\end{eqnarray}
The index $HYBX$ refers to hybrid-X, where $X$ can indicate the value of $x$: e.g. two candidates are $\widehat{V} _{HYB01}(\hat{\theta}) $  and $\widehat{V} _{HYB05}(\hat{\theta}) $  for $x=0.01$ and $x=0.05$ respectively.

\section{Discussion of practical determination of input parameters}
\label{prac_det}

\subsection{Determination of $M_{sample}$ and $\hat{\theta}$ }
It is generally possible to determine the sample mass $M_{sample}$ accurately using weighing. The sample concentration $\hat{\theta}$ is generally determined using chemical, physical or biological testing in a laboratory as determining this value is generally part of sampling and estimation of the concentration in the population. It can therefore be assumed that a numerical value for $\hat{\theta}$  is known.

\subsection{Determination of $N_i$}
For small samples it might be feasible to classify particles one by one by hand and to establish $N_i$ for each class by counting. However, practical samples may contain thousands of small particles, which may also be difficult to classify visually. Indirect or automated methods to determine $N_i$ need to be applied in these cases. It might also be possible to evaluate the variance estimators without numerical value of $N_i$: all estimators, except $\widehat{V}_{SYG}$, depend on $N_i$ only as the product $N_i m_i$ which is equal to the total mass of material belonging to the $i$th particle class in the sample. If the materials belonging to the separate classes in the sample can be separated $N_i m_i$ can be determined by weighing directly. 

\subsection{Determination of $m_i$ and $c_i$}
Within a class, particle mass ($m_i$) and concentration ($c_i$) are constant, so in principle determining the mass of one particle by weighing and concentration by chemical, physical or biological testing would suffice to establish the particle properties of the entire class. However, it is recommended here to analyze more than one particle to assert $m_i$ and $c_i$ are constant within a particle class. It may also be infeasible to analyze a single particle if it is too small.  In some cases, there may be prior knowledge about the material types. In those cases, $c_i$ can possibly be estimated using the known material properties. The particle mass can sometimes be determined using the product of particle volume and material density. 

\subsection{Determination of $C_{ij}$}
It has been discussed by \cite{r2} that a negative value of $C_{ij}$ implies grouping/clustering of particles, while a positive value of $C_{ij}$ implies segregation of particles. Currently, research is being conducted to evaluate $C_{ij}$ using image analysis and/or a modeling approach of particle properties. First principles of such an image-based approach to determine the value of $C_{ij}$ have recently been established \cite{r9}.

\section{Conclusion}
Six variance estimators for use in the application area of particulate material sampling were constructed.

\section{Acknowledgements}
This work was performed as part of a project supported by the Netherlands Technology Foundation STW, under STW grant 7457. NFI, Deltares, Nutreco, Hosokawa and Organon are members of the users' committee of this project.

\appendix
\section{The parameter for the dependent selection of particles}\label{app}
If $\pi _i$  and $\pi _{ij}$  are respectively the first and second order inclusion probabilities of the $i$th and $i$th and $j$th particle of the population and $n(i)$ is the class number of the $i$th particle and $n(j)$ is the class number of the $j$th particle, then we consider the class of sampling designs for which:
\begin{equation}
\label{kappa_i}
\kappa _{n(i)} \equiv \pi _i
\end{equation}
\begin{equation}
\label{kappa_ij}
\kappa _{n(i) n(j)} \equiv \pi _{ij}
\end{equation}
Because each particle belonging to an arbitrary class $i$, has a probability  $\kappa _i$ of being included in the sample, the expected number of particles belonging to the $i$th class in the sample is equal to:
\begin{equation}
E(N_i)=\kappa _i N_{i,batch}	
\end{equation}
where $N_{i,batch}$ is the number of particles belonging to the $i$th class in the population. From the above equation follows that $\kappa _i=E(N_i)/N_{i,batch}$. Similarly to the above derivation, an expression for the second-order inclusion probability $\kappa _{ij}$ can be derived. Note that only when the two particles $i$ and $j$ are selected independently $\kappa _{ij}= \kappa _i \times \kappa _j$. In all other cases, a correction factor is required \cite{r1}. This is written as:
\begin{equation}
\kappa _{ij}= \kappa _i \times \kappa _j \times (1-{C}' _{ij})
\end{equation}
where ${C}'_{ij}$ is the 'parameter for the dependent selection of particles'. It will now be proven that ${C}'_{ij} = C_{ij}$. A population with $N_{i,batch}$ and $N_{j,batch}$ particles belonging to the $i$th and $j$th class ($i \neq j$) contains $N_{i,batch} \times N_{j,batch}$ pairs of particles where the first particle is of type $i$ and the second of type $j$. If $i=j$, there are  $N_{i,batch} \times (N_{j,batch}-1)$ such pairs. For the sample, the numbers of pairs follow a similar pattern: $N_i \times N_j$ pairs if $i \neq j$ and $N_i \times (N_j-1)$ if $i=j$. Because the expected number of pairs in the sample is equal to the number of pairs in the population multiplied by the probability of a pair of being included in the sample, the expected value of the number of pairs is written as:
\begin{equation}
E(N_i \times (N_j - \Delta _{ij})) =
N_{i,batch} \times (N_{j,batch} - \Delta _{ij}) \times \kappa _{ij}
\end{equation}
Where $\Delta _{ij}$ is the Kronecker delta, a parameter whose value is one if $i=j$ and zero otherwise. The above equations, combined with Eq. (\ref{Cij}), can be used to obtain the following expression:
\begin{equation}
C_{ij} = {C}' _{ij} + \Delta _{ij} (1 - {C}' _{ij})/N_{i,batch}
\end{equation}
In the limit of infinite numbers of particles within each class of particles in the population (or batch) $C _{ij} = {C}' _{ij}$. This proves Eq. (\ref{Cij_inc}) in the main text.


\begin{thebibliography}{}

\bibitem{r1}
Geelhoed B (2004), Sampling of particulate materials - New theoretical approach, 200 pp, Delft University Press, Delft (ISBN 90-407-2517-9)

\bibitem{r2}
Geelhoed B (2005), A Sampling Study of Industrial Mixtures of Particles, p. 48-50 in:  Stare J, Vidmar, G, Koren G (eds.) (2005) Applied Statistics 2005, PROGRAM and ABSTRACTS, 98 pp,
Statistical Society of Slovenia, Ljubljana  (ISBN 961-90314-4-X)

\bibitem{r3}
Geelhoed B (2006),
Variable second-order inclusion probabilities during the sampling of industrial mixtures of particles,
Applied Stochastic Models in Business and Industry 22: 495-501 

\bibitem{r5}
S\"{a}rndal C E, Swensson B, Wretman J (1992),
Model Assisted Survey Sampling, Springer-Verlag, New York

\bibitem{r6}
Sen A (1953), On the estimate of the variance in sampling with varying probabilities,
Journal of Indian Society for Agricultural Statistics 5: 119-127

\bibitem{r7}
Yates F, Grundy P (1953),
Selection without replacement from within strata with probability proportional to size,
Journal of the Royal Statistical Society B 15: 235-261

\bibitem{r9}
Dihalu D, Geelhoed B (2009),
Principles of an image-based algorithm for the quantification of dependencies between particle selections in sampling studies, Fourth World Conference on Sampling and Blending, The Southern African Institute of Mining and Metallurgy S59: 243-250

\bibitem{r12}
Cleary PW, Robinson GK, Golding MJ, Owen PJ (2008),
Understanding factors leading to bias for falling-stream cutters using discrete element modelling with
non-spherical particles,
Chemical Engineering Science 63: 5681-5695 


\end{thebibliography}
\end{document}